\documentclass[11pt,reqno,twoside]{amsart}
\usepackage{geometry}
\usepackage{stmaryrd}
\usepackage[pagewise]{lineno} 
\usepackage{amssymb}
\usepackage[dvips]{epsfig}
\usepackage{graphicx}
\usepackage{color}
\usepackage{amsmath}
\usepackage{amssymb}
\usepackage{amsfonts}
\usepackage{amsthm}
\usepackage{mathrsfs}
\newcommand{\im}{\mathbf{i}}

\newcommand{\R}{\mathbb R}

\newcommand{\Z}{\mathbb Z}

\newcommand{\ep}{\varepsilon}

\renewcommand{\lg}{\Lambda}

\renewcommand{\u}{\mathscr{U}}

\newcommand{\T}{\mathbb{T}}

\newcommand{\boA}{\mathcal{A}}
\newcommand{\boB}{\mathcal{B}}
\newcommand{\boD}{\mathcal{D}}

\newcommand{\boG}{\mathcal{G}}

\newcommand{\boI}{\mathcal{I}}

\newcommand{\boP}{\mathcal{P}}

\newcommand{\sA}{\mathscr{A}}
\newcommand{\sB}{\mathscr{B}}
\newcommand{\sC}{\mathscr{C}}
\newcommand{\sD}{\mathscr{D}}

\newcommand{\sU}{\mathscr{U}}
\newcommand{\sW}{\mathscr{W}}
\newcommand{\sG}{\mathscr{G}}
\newcommand{\sT}{\mathscr{T}}
\newcommand{\sI}{\mathscr{I}}

\newcommand{\sH}{\mathscr{H}}
\newcommand{\sP}{\mathscr{P}}

\newcommand{\sM}{\mathscr{M}}
\newcommand{\sX}{\mathscr{X}}
\newcommand{\sY}{\mathscr{Y}}

\newcommand{\sL}{\mathscr{L}}
\newcommand{\sK}{\mathscr{K}}
\newcommand{\sR}{\mathscr{R}}
\newcommand{\sQ}{\mathscr{Q}}
\newcommand{\sV}{\mathscr{V}}

\newcommand{\ji}{\langle}
\newcommand{\jd}{\rangle}

\newcommand{\dist}{{\rm dist}}

\newtheorem{thm}{Theorem}[section]
\newtheorem{lem}[thm]{Lemma}

\newtheorem{prop}[thm]{Proposition}
\newtheorem{cor}[thm]{Corollary}
\theoremstyle{remark}
\newtheorem{rem}{\bf Remark}[section]
\theoremstyle{definition}

\numberwithin{equation}{section}

\usepackage[colorlinks=true,pdfstartview=Fit,linkcolor=magenta,citecolor=cyan]{hyperref}
\usepackage{bm}
\DeclareMathOperator{\arsinh}{arsinh}
\DeclareMathOperator{\Tr}{Tr}
\DeclareMathOperator{\mes}{mes}
\DeclareMathOperator{\Var}{Var}
\begin{document}

\title[All coupling localization in QKR]{All coupling Anderson localization in the quantum kicked rotor model}
	\author[Shi]{Yunfeng Shi}
\address[Shi] {School of Mathematics,
	Sichuan University,
	Chengdu 610064,
	China}
\email{yunfengshi@scu.edu.cn}

\author[Wen]{Li Wen}
\address[Wen] {School of Mathematics,
	Sichuan University,
	Chengdu 610064,
	China}
\email{liwen.carol98@gmail.com}
\date{\today}
	\keywords{Anderson localization; Quantum kicked rotor; Cayley transform; Weyl sum estimate}

\begin{abstract}
In this paper, we study  $$\sH=\lambda \Delta +\tan\pi(\theta+n^2\alpha)\delta_{n,n'},\ n\in\Z$$ and prove its Anderson localization for all $\lambda >0$ and Lebesgue almost every (a.e.)  $(\alpha, \theta)\in\T^2$. This conforms in the full measure sense  the main assumption of Fishman-Grempel-Prange [{\it Phys. Rev. Lett.}  49(8), 509--512 (1982)]  in  the study of  the  quantum suppression of chaos: The pseudo-randomness  of  the sequence $\{\tan\pi(\theta+n^2\alpha)\}_{n\in\mathbb Z}$ (for $\alpha\in\T\setminus \mathbb Q$)  should result  in all coupling Anderson  localization for  $H$ similar to that  for  the one-dimensional Anderson model.  Our proof is based on nonperturbative Green's function estimates approach using Cayley transform, quadratic Weyl sum estimate and Rellich function analysis.
\end{abstract}

\maketitle

\section{Introduction and Main result}\label{sec:introduction}

One of the main models  in quantum chaos is the so-called  $1$-dimensional quantum kicked rotor  which is given by the Floquet Hamiltonian 
\begin{align}\label{qkrm}
	\sqrt{-1}\frac{\partial \Psi(x, t)}{\partial t}= \left(\frac{\partial^2 }{\partial x^2}+\lambda \cos(2\pi x)\sum_{n\in\Z}\delta(t+2n \alpha)\right)\Psi(x, t),
\end{align}
where $(x, t)\in \T\times \R, \ \alpha\in\T \setminus\mathbb Q$ and $\lambda\in\R$. 
This  model  was first introduced by Casati {et. al.} \cite{Cas79} as a quantum analogue of the usual Chirikov standard map.  It was conjectured  by  the remarkable work \cite{FGP82} of Fishman, Grempel and Prange     that  the  motion in  the quantum kicked rotor is generally  almost-periodic  even though  it is  typically chaotic in the classical one {\bf for large $\lambda$}.  This is the famous  {\bf quantum suppression of chaos} phenomenon.   Indeed, in \cite{FGP82},  by introducing an elegant transformation,  the authors  reduced the quantum kicked rotor  to a  discrete operator $\sH$  of the form
\begin{align*}
	(\sH u)_n=\sum_{m\in\Z}\phi_{n-m}u_m+\tan\pi\left(\theta+n^2\alpha \right)u_n,\ n\in\Z,\  \theta\in\R,
\end{align*}
where
$\phi_n=-\int_{\T} \tan\pi\left(\frac{\lambda\cos (2\pi x)}{2}\right)e^{-2\pi\sqrt{-1} nx}\mathrm{d}x.$
Under this  reduction,  they further argued that the occurrence of  the Anderson localization (i.e., pure point spectrum with exponentially decaying eigenfunctions) for $\sH$ will  imply the almost-periodicity of solutions for the quantum kicked rotor \eqref{qkrm}, and then the absence of chaos (cf. \cite{FGP82,SPK22}).  The  quantum suppression of chaos problem   could eventually be  solved   \cite{FGP82} under the {\bf main assumption}  (supported by the numerical evidence) that the {\bf pseudo-randomness}  of  the sequence $\{\tan\pi(\theta+n^2\alpha)\}_{n\in\mathbb Z}$ (for $\alpha\in\T\setminus \mathbb Q$)  should result  in {\bf all coupling Anderson  localization}  for  $\sH$ similar to that  for  the one-dimensional Anderson model \cite{And58}. 

In the present work, we aim to conform in the full measure sense  this  main assumption and proved {\bf all coupling Anderson localization}  for the nearest-neighbor version of $\sH$ (also suggested by Fishman et al. \cite{FGP82}).

The popular mathematical  models    in the study of quantum kicked rotor problems  are  typically  related to  lattice  Schr\"odinger operators with potentials given by the skew-shift: 
$$\sH=\lambda \Delta +v(\sT_\alpha^n  (\theta_1, \theta_2))\delta_{n,n'},\ n\in\Z,$$ 
where $\Delta$ is the discrete Laplacian\footnote{One also considers  more general long-range hopping operators \cite{Bou02, Bou05}.},  $v$ is a real function\footnote{One also considers  $v$ on $\T$  with $\sT_\alpha^n (\theta_1, \theta_2)=\theta_1+n \theta_2+\frac{n(n-1)\alpha}{2}$ \cite{Bou02}. } on $\T^2$ and $\sT (\theta_1, \theta_2)=(\theta_1+\theta_2, \theta_2+\alpha).$
Bourgain-Goldstein-Schlag \cite{BGS01}  first proved  the Anderson localization for $H$ for non-constant analytic $v$ and small enough  $\lambda$ provided $\alpha$ belonging to  a  set of  large Lebesgue measure. The strategy is based on large deviation estimates on Green's function  and Lyapunov exponent analysis in the sense of  \cite {BG00, GS01}.  Based on multi-scale analysis type method related to \cite{BGS02}, Bourgain \cite{Bou02}  treated the  more general  skew-shift operators with  Jacobi type hopping and established Anderson localization if the hopping strength is small enough. This leads to the resolution the  problem of \cite{Sin, Bel}  that,  for typical quantum kicked rotor models with small  potentials, its  wave function will be almost-periodic in time \cite{Bou02}.   Recently, Shi-Yuan \cite{SY} modified the method of  \cite{BGS01, Bou02} to prove the Anderson  localization for skew-shift operators   with the tangent potential   assuming $\lambda$ is small. 

All the above localization results need the  $\lambda$ to be small enough.  However, for the cosine skew-shift operators,  it was conjectured by Bourgain that {\bf all coupling  Anderson localization}  should be expected (cf. e.g., Chapter 15, \cite{Bou05}). For the cosine skew-shift model at a particular phase, Bourgain \cite{Bou02b} has made a major breakthrough and proved the existence of point spectrum whose closure has positive measure  for all $\lambda\neq 0$ and  for $\alpha$ on a set of  positive  measure.  To the best of our knowledge,  the present work provides the first result on the  all coupling Anderson localization  for a skew-shift type self-adjoint Schr\"odinger operator.

Our proof differs from the approaches of Bourgain-Goldstein-Schlag \cite{BGS01} and Bourgain \cite{Bou02}. It relies on nonperturbative Green’s function estimates and is closer in spirit to the work of Jitomirskaya \cite{Jit94,Jit99}. More precisely, we exploit the special structure of the tangent potential to obtain, via the Cayley transform, a three-part decomposition of the Dirichlet determinants. One of our main contributions is to establish quantitative equidistribution estimates for the zeros of the relevant determinants using bounds on quadratic Weyl sums. These estimates lead to large-deviation estimates for Green’s functions for every coupling strength ($\lambda\neq 0$). Finally, we eliminate double resonances through a transversality analysis of Rellich functions. Combining the resulting Green’s function estimates with Schnol’s lemma then yields Anderson localization.

\subsection{Main result}
We study the following skew-shift Schr\"odinger operators
\begin{align}\label{eq:quadratic-tangent-operator}
	\sH_{\lambda,\theta,\alpha}=\lambda\Delta+\tan\pi(\theta+n^2\alpha)\delta_{n,n'},\ n,n'\in\Z,
\end{align}
where $\lambda>0$, $\theta,\alpha\in\T:=\R/\Z$ and the discrete Laplacian $\Delta$ is defined as 
\begin{align*}
	\Delta(n,n')=\delta_{|n-n'|,1}.
\end{align*}
Let ${\rm mes}(\cdot)$ denote the Lebesgue measure. Define  the set of poles by
\begin{equation}\label{eq:tangent-pole-set}
\mathcal{P}=\bigcup_{n\in\mathbb{Z}}
\left\{(\theta,\alpha)\in\mathbb{T}^2:\ 
\theta+n^2\alpha\equiv\tfrac12\pmod{1}\right\}.
\end{equation}
 Then we have
\begin{thm}\label{thm:all-coupling-localization}
For fixed $\lambda>0$, there is $\mathcal{G}_\lambda\subset\mathbb{T}^2\setminus\mathcal{P}$ with $\mes\mathcal{G}_{\lambda}=1$ such that, for every $(\theta,\alpha)\in\mathcal{G}_\lambda$, the operator \eqref{eq:quadratic-tangent-operator} has pure point spectrum. Moreover, for each eigenvalue $E$ and corresponding eigenfunction $\psi$, there is a constant $C=C(\lambda,E,\psi,\theta,\alpha)>0$ such that
\begin{equation}\label{eq:eigenfunction-exponential-decay}
|\psi(n)|\leq Ce^{-m_\lambda|n|},
\ m_\lambda=\frac1{40}\operatorname{arsinh}\frac1{2\lambda}>0,
\end{equation}
where $\arsinh x=\log(x+\sqrt{x^2+1})$ for $x\in\R$.
\end{thm}

\subsection{Idea of the proof and new ingredients}

Clearing the tangent denominators gives the Cayley factorization
\begin{align*}
	\det(\sH_N-E\sI_N)=A_N(E)\frac{Q_N(w;E)}{B_N(w)},\ w=e^{2\pi\im\theta}.
\end{align*}
The roots of $B_N$ are explicit, and quadratic Weyl sums control how
evenly they are distributed on the unit circle. For $Q_N$, we
truncate the free Cayley kernel and approximate the unitary trace
$\Tr(\sU_{E,N}^k)$ by a sum over closed paths. The phase along each
path has quadratic coefficient $k\alpha$ in the starting point.
Since the Weyl sum estimate is uniform, it controls the distribution
of the roots of both $B_N$ and $Q_N$.

We also show that a root of $Q_N$ close to the phase $w$ would produce
an approximate eigenvector at energy $E$. Thus, if $E$ stays away
from the finite-volume spectrum, then $w$ must stay away from every
root of $Q_N$. By working with the matrix obtained after clearing
the tangent denominators, this argument also covers roots at tangent poles.

In the cofactor formula, the $B$ factors cancel outside the interval
joining the two points. The remaining $B$ and $Q$ factors contribute
sublinear logarithmic errors, and the growth of $A_N$ gives
exponential Green function decay on nonresonant intervals.
\subsection{Structure of the paper}
The paper is organized as follows.
After collecting the notation, we derive the Cayley factorization
and estimate $A_N$ in Section~\ref{sec:cayley-factorization}.
Section~\ref{sec:phase-and-spectral-factors} establishes bounds for
$B_N$ and $Q_N$. Section~\ref{sec:green-function-decay} combines the
three factor estimates to obtain Green function decay on nonresonant
intervals. In Section~\ref{sec:spectral-separation-and-localization},
we exclude double resonances and complete the proof of
Theorem~\ref{thm:all-coupling-localization}.

Appendix~\ref{app:determinant-factor-estimates} collects auxiliary
estimates for the determinant factors.
Appendix~\ref{app:phase-root-and-block-spectral-separation} gives
the details of phase-root separation and the pole count in the
shear argument.
\section{Notation}
We collect the notation used throughout the paper.
\begin{itemize}
	\item Let $\im=\sqrt{-1}$ be the imaginary unit.
	
	\item For $x\in\R$, $\arsinh x=\log(x+\sqrt{x^2+1})$.
	\item The determinant of a matrix $\sM$ is denoted by $\det \sM$.
	
	\item For $\boA,\boB\subset\R$, we define
	\begin{align*}
		\dist(\boA,\boB)=\inf\{|a-b|:\ a\in\boA,\ b\in\boB\}.
	\end{align*}
	
		\item Let $\bm 1_{\boA}$ be the indicator function of set $\boA$.
	
	\item Let $\#\mathcal{A}$ denote the cardinality of set $\mathcal{A}$.
	
	\item For $x\in\R$, we define $\|x\|_{\T}=\inf_{l\in\Z}|l-x|$.
	
	\item $\{\delta_{x}\}_{x\in\Z}$ is the standard basis of $\ell^2(\Z)$.
	
	\item $\sI$ typically denotes the identity operator.
	
	\item For a linear operator $\sT$, let $\|\sT\|$, $\|\sT\|_{\mathfrak{S}_1}$, ${\rm rank}(\sT)$ and $\Tr(\sT)$ denote the operator norm, the trace norm, the rank and the trace of $\sT$, respectively, 
	
	\item $\sP_{\lg}$ denotes the orthogonal projection onto
	$\ell^2(\lg)$, where $\lg\subset\mathbb Z$.
	
	\item Denote by $\ji\cdot,\cdot\jd$ the standard inner product on $\ell^2(\Z)$.
	
	\item Let $\sT:\ \ell^2(\Z)\to\ell^2(\Z)$ be a linear operator.  Set $\sT(x,y)=\ji \delta_{x}, \sT\delta_{y}\jd$. The spectrum of $\sT$ is denoted by $\sigma(\sT)$. Finally, we define $\sT_\lg=\sP_\lg \sT\sP_\lg$, where $\lg\subset\Z$. In particular, let $\lg_{N}=[0,N-1]\cap\Z$ and we simplify $\sT_{\lg_N}$ to $\sT_{N}$.
	
	\item For $E\notin\sigma(\sH_{\boI})$, we define its Green's function by
	\begin{align*}
		\sG_{\boI}(x,y;E)=\ji \delta_{x}, (\sH_{\boI}-E\sI_{\boI})^{-1}\delta_{y}\jd,\ x,y\in\boI.
	\end{align*}
	
	\item For a finite Borel measure $\mu$ on $\T$, its Fourier coefficients are
	\begin{align*}
		\hat{\mu}(k)=\int_{\T}e^{2\pi\im kt}d\mu(t),\ k\in\Z
	\end{align*}
	and for $f\in L^1(\T)$ we use 
	\begin{align*}
		\hat{f}(k)=\int_{\T}f(t)e^{-2\pi\im kt}dt,\ k\in\Z.
	\end{align*}
	\item  We denote by $BV(\mathbb T)$ the space of functions $f\in L^1(\T)$ with finite total variation.
	For $f\in BV(\mathbb T)$, we write $\Var_{\mathbb T}(f)$ for the total variation.
\end{itemize}

\section{Cayley factorization and the free determinant}\label{sec:cayley-factorization}

\subsection{Separating the three factors}
We first remove the tangent denominators and identify the three factors
that will be estimated below. On
$\Lambda_N=\{0,\ldots,N-1\}$, use the phase
\begin{align}\label{eq:local-quadratic-phase}
q_j=\theta+\beta j+\alpha j^2.
\end{align}
Locally, the diagonal entries of $\sH_N$ are given by $\tan\pi q_j$.
For an interval starting at $b\in\mathbb Z$, we write its points as $b+j$. Then
\begin{align*}
	\theta+\alpha(b+j)^2=\theta_b+\beta_b j+\alpha j^2,
\end{align*}
where
\begin{align}\label{eq:translated-phase-parameters}
	\theta_b=\theta+b^2\alpha,\ \beta_b=2b\alpha.
\end{align}

For $E\in\R$, define
\begin{equation}\label{eq:free-matrix-and-phase-matrix}
\sK_{E,N}=\lambda\Delta_N-E\sI_{N},
\ \sD_N=\sD_{\alpha,\beta,N}
=\operatorname{diag}_{0\le j\le N-1}\left(e^{2\pi\im(\alpha j^2+\beta j)}\right),
\end{equation}
and
\begin{equation}\label{eq:finite-cayley-matrices}
\sC_{E,N}=(\sK_{E,N}+\im\sI_{N})^{-1}(\sK_{E,N}-\im\sI_{N}),
\ \sU_{E,N}=\sC_{E,N}\sD_N.
\end{equation}
Since $\sK_{E,N}$ is self-adjoint, both $\sC_{E,N}$ and $\sU_{E,N}$ are unitary.
Set $w=e^{2\pi\im\theta}$ and write
\begin{align*}
	A_N(E)&=\det(\sK_{E,N}+\im\sI_{N}),\\
	B_N(w)&=\det(\sI_{N}+w\sD_N),\\
	Q_N(w;E)&=\det(\sI_{N}+w\sU_{E,N}).
\end{align*}
Here $A_N$ is the free determinant, $B_N$ is the product of the
tangent denominators, and the roots of $Q_N$ are determined by the
eigenvalues of $\sU_{E,N}$.

To compare determinants on different subintervals, we keep the original integer points. For an interval $\mathcal{I}\subset\Z$, define
\begin{align*}
	\sK_{E,\mathcal{I}}&=\lambda\Delta_{\mathcal{I}}-E\sI_{\boI},\\
	\sD_{\mathcal{I}}&=\operatorname{diag}_{n\in \mathcal{I}}\left(e^{2\pi\im\alpha n^2}\right),\\
	\sC_{E,\mathcal{I}}&=(\sK_{E,\mathcal{I}}+\im\sI_{\boI})^{-1}(\sK_{E,\mathcal{I}}-\im\sI_{\boI}),\\
	\sU_{E,\mathcal{I}}&=\sC_{E,\mathcal{I}}\sD_{\mathcal{I}}.
\end{align*}
We then set
\begin{align*}
		A_{\mathcal{I}}(E)&=\det(\sK_{E,\mathcal{I}}+\im\sI_{\boI}),\\
	B_{\mathcal{I}}(w)&=\det(\sI_{\boI}+w\sD_{\mathcal{I}}),\\
	Q_{\mathcal{I}}(w;E)&=\det(\sI_{\boI}+w\sU_{E,\mathcal{I}}).
\end{align*}
Thus all subintervals use the same phase variable $w=e^{2\pi\im\theta}$.
We use the change of coordinates \eqref{eq:translated-phase-parameters} only when applying
estimates stated on $\Lambda_N$.

\begin{lem}\label{lem:cayley-determinant-factorization}
Let $w=e^{2\pi\im\theta}$. If $q_j\not\equiv\tfrac12\pmod1$ for every $j\in \lg_{N}$, where $q_j$ is defined by \eqref{eq:local-quadratic-phase}, then
\begin{equation}\label{eq:denominator-cleared-factorization}
\begin{aligned}
(\sH_N-E\sI_{N})(\sI_{N}+w\sD_N)
&=(\sK_{E,N}+\im\sI_{N})+(\sK_{E,N}-\im \sI_N)w\sD_N\\
&=(\sK_{E,N}+\im \sI_N)(\sI_{N}+w\sU_{E,N}).
\end{aligned}
\end{equation}
Hence
\begin{equation}\label{eq:three-factor-determinant}
\det(\sH_N-E\sI_{N})=A_N(E)\frac{Q_N(w;E)}{B_N(w)}.
\end{equation}
The polynomial $Q_N(\cdot;E)$ has degree $N$, and all its roots lie on the unit circle.
\end{lem}

\begin{proof}
	Using the identity
	\begin{align*}
		\tan\pi q=\im\frac{1-e^{2\pi\im q}}{1+e^{2\pi\im q}},
	\end{align*}
	we obtain
	\begin{align*}
			\sV(\sI_N+w\sD_N)
		=\im(\sI_N-w\sD_N),
	\end{align*}
	where $\sV=\operatorname{diag}_{j\in\lg_N}(\tan\pi q_j)$. From the fact that
	\begin{align*}
		\sH_N-E\sI_{N}=\sK_{N}+\sV,
	\end{align*}
	we have
	\begin{align*}
		(\sH_N-E\sI_{N})(\sI_N+w\sD_N)&=(\sK_N+\im \sI_N)+\sV(\sI_N+w\sD_N)+w\sK_N\sD_N-\im\sI_N\\
		&=(\sK_N+\im \sI_N)+(\sK_N-\im\sI_N)w\sD_N\\
		&=(\sK_N+\im \sI_N)\left(\sI_N+(\sK_N+\im \sI_N)^{-1}(\sK_N-\im\sI_N)w\sD_N\right)\\
		&=(\sK_{E,N}+\im \sI_N)(\sI_{N}+w\sU_{E,N}).
	\end{align*}
	 Taking determinants then yields \eqref{eq:three-factor-determinant}. If $u_1,\cdots,u_N$ are the eigenvalues of $\sU_{E,N}$, then
	\begin{align*}
		Q_N(w;E)=\prod_{j=1}^N(1+wu_j),\ |u_j|=1.
	\end{align*}
Its roots are therefore $-u_j^{-1}$, and its leading coefficient $\det\sU_{E,N}$ is nonzero. We have finished the proof.
\end{proof}

\begin{rem}\label{rem:phase-roots-at-poles}
At a pole-free phase, a zero of $Q_N(\cdot;E)$ corresponds to an eigenvalue $E$ of $\sH_N$. Even at a tangent pole, the matrix polynomial in \eqref{eq:denominator-cleared-factorization} remains well-defined. We will use its kernel in Lemma~\ref{lem:spectral-gap-to-phase-root-separation} to compare the corresponding root with nearby pole-free phases.
\end{rem}

\subsection[Exponential growth of A]{Exponential growth of \texorpdfstring{$A_N$}{A}}\label{subsec:free-determinant-growth}

We begin with the phase-independent factor. Its explicit recurrence
gives the positive exponent that will survive in the cofactor quotient.

For $E\in\R$, let $\rho_E$ denote the unique solution of
\begin{equation}\label{eq:free-characteristic-equation}
	\rho+\rho^{-1}=\frac{E-\im}{\lambda}
\end{equation}
satisfying $|\rho_E|>1$, and define
\begin{align*}
	g(E,\lambda)=\log|\rho_E|.
\end{align*}
Set
\begin{align*}
	g_\lambda=\operatorname{arsinh}\frac1{2\lambda},
	\ 
	J(\lambda)=\log\frac{2}{1-e^{-2g_\lambda}}.
\end{align*}
\begin{lem}\label{lem:free-determinant-growth}
For $n\geq0$, we have
\begin{equation}\label{eq:free-determinant-formula}
A_n(E)=(-\lambda\rho_E)^n\frac{1-\rho_E^{-2n-2}}{1-\rho_E^{-2}}.
\end{equation}
Moreover,
\begin{equation}\label{eq:uniform-free-growth-rate}
g(E,\lambda)\geq g_\lambda>0
\end{equation}
and
\begin{equation}\label{eq:free-determinant-log-estimate}
\left|\log|A_n(E)|-n\big(\log\lambda+g(E,\lambda)\big)\right|
\leq J(\lambda).
\end{equation}
\end{lem}

\begin{proof}
Expanding the tridiagonal determinant along the last row gives
\begin{align*}
	A_n=(-E+\im)A_{n-1}
	-\lambda^2 A_{n-2},\
	A_0=1,\ A_1=-E+\im.
\end{align*}
The two corresponding exponential solutions are $(-\lambda\rho_E)^n$ and $(-\lambda\rho_E^{-1})^n$. Formula~\eqref{eq:free-determinant-formula} satisfies the recurrence and the initial conditions. Writing $\rho_E=e^{g+\im\varphi}$, we obtain from \eqref{eq:free-characteristic-equation}
\begin{align*}
	\lambda^{-1}=2\sinh g\,|\sin\varphi|
	\leq 2\sinh g,
\end{align*}
which yields \eqref{eq:uniform-free-growth-rate}. Moreover, we have
\begin{align*}
	1-e^{-2g_\lambda}\le |1-\rho_E^{-2n-2}|,|1-\rho_E^{-2}|\le 2.
\end{align*}
Consequently, we can get
\begin{align*}
	\left|
	\log\frac{|1-\rho_E^{-2n-2}|}
	{|1-\rho_E^{-2}|}
	\right|
	\leq \log\frac{2}{1-e^{-2g_\lambda}},
\end{align*}
which gives \eqref{eq:free-determinant-log-estimate}. This completes the proof.
\end{proof}

\section[The factors B and Q]{The factors \texorpdfstring{$B_N$}{B} and \texorpdfstring{$Q_N$}{Q}}\label{sec:phase-and-spectral-factors}

The free determinant supplies the linear term in the logarithmic
Green function bound. We now show that the two factors contribute
smaller errors, starting with the explicit roots of $B_N$ and then
passing to the spectral roots of $Q_N$.

\subsection[Quadratic phases and the product B]
{Quadratic phases and the product \texorpdfstring{$B_N$}{B}}
\label{subsec:quadratic-phase-discrepancy}

The eigenvalue angles of $\sD_N$ have normalized counting measure
\begin{align*}
\mu_{D,N}(\boA)=\frac1N\sum_{n=0}^{N-1}\bm 1_{\boA}(\alpha n^2+\beta n)\ \text{for $\boA\subset\T$}.
\end{align*}
The roots of $B_N$ have angles $\tfrac12-\alpha n^2-\beta n$.
For a probability measure $\mu$ on $\T$, define its discrepancy by
\begin{align*}
\mathfrak D(\mu)=\sup_{\mathcal J}|\mu(\mathcal J)-|\mathcal J||,
\end{align*}
where the supremum is over half-open arcs. Reflection and translation
preserve discrepancy, so it suffices to estimate $\mu_{D,N}$.
The Erd\H{o}s-Tur\'an inequality \cite{ET48} gives, for every integer $K\geq1$,
\begin{equation}\label{eq:erdos-turan-discrepancy}
\mathfrak D(\mu)\leq\frac1{K+1}
+2\sum_{k=1}^K\frac{|\hat\mu(k)|}{k}.
\end{equation}
In the present case,
\begin{align*}
\hat\mu_{D,N}(k)=\frac1N\sum_{n=0}^{N-1}
e^{2\pi\im(k\alpha n^2+k\beta n)},
\end{align*}
so we first need a quadratic Weyl bound uniform in the linear phase.

For $0<\gamma<1/2$, define
\begin{equation}\label{eq:diophantine-frequency-set}
\boD_\gamma=
\left\{\alpha\in\T:
\|k\alpha\|_{\T}\geq\gamma|k|^{-2}
\text{ for all }k\in\Z\setminus\{0\}\right\}.
\end{equation}
For each $k\geq1$, the exceptional set in \eqref{eq:diophantine-frequency-set} has measure $2\gamma k^{-2}$. Hence, by the union bound, we obtain
\begin{equation}\label{eq:diophantine-exceptional-measure}
	\mes(\mathbb T\setminus\boD_\gamma)
	\leq 2\gamma\sum_{k\geq1}k^{-2}
	=\frac{\pi^2}{3}\gamma.
\end{equation}
We will use the lower bound in \eqref{eq:diophantine-frequency-set} after taking differences
of the quadratic phase. The following inequality controls the sum through its shifted correlations.

\begin{lem}\label{lem:van-der-corput-differencing}
If $|z_n|\leq1$ for $0\leq n<N$ and $1\leq H\leq N$, then
\begin{equation}\label{eq:van-der-corput-bound}
\left|\frac1N\sum_{n=0}^{N-1}z_n\right|^2
\leq\frac{2}{H}+\frac{4}{NH}
\sum_{h=1}^{H-1}
\left|\sum_{n=0}^{N-h-1}z_{n+h}\overline{z_n}\right|.
\end{equation}
\end{lem}

\noindent The proof is given in Appendix~\ref{app:determinant-factor-estimates}.

\begin{lem}\label{lem:uniform-quadratic-weyl-sums}
Let $W(\gamma)=\sqrt{4+8/(3\gamma)}$.
If $\alpha\in\boD_\gamma$, then, for every $N\geq1$ and
$1\leq k\leq N^{1/16}$,
\begin{equation}\label{eq:quadratic-weyl-sum-bound}
\sup_{b\in\R}
\left|\frac1N\sum_{n=0}^{N-1}
e^{2\pi\im(k\alpha n^2+bn)}\right|
\leq W(\gamma)N^{-1/8}.
\end{equation}
\end{lem}

\begin{proof}
Set $z_n=e^{2\pi\im(k\alpha n^2+bn)}$. Then we have
\begin{align*}
z_{n+h}\overline{z_n}
=e^{2\pi\im(k\alpha h^2+bh)}
 e^{2\pi\im\,2kh\alpha n}.
\end{align*}
The geometric sum bound and \eqref{eq:diophantine-frequency-set} therefore give
\begin{align*}
\left|\sum_{n=0}^{N-h-1}z_{n+h}\overline{z_n}\right|
&\leq \min\left\{N,\frac1{2\|2kh\alpha\|_{\T}}\right\}
\leq2\gamma^{-1}k^2h^2.
\end{align*}
Substituting into \eqref{eq:van-der-corput-bound}, we obtain
\begin{align*}
	\left|\frac1N\sum_{n=0}^{N-1}z_n\right|^2
	\leq\frac2H+\frac{8k^2H^2}{3\gamma N}.
\end{align*}
Taking $H=\lfloor N^{1/4}\rfloor$ and using
$k\leq N^{1/16}$ gives
\begin{align*}
	\left|\frac1N\sum_{n=0}^{N-1}z_n\right|^2
	\leq\left(4+\frac8{3\gamma}\right)N^{-1/4}.
\end{align*}
Taking square roots proves \eqref{eq:quadratic-weyl-sum-bound}. The proof has been finished.
\end{proof}

For $\alpha\in\boD_\gamma$, taking $K=\lfloor N^{1/16}\rfloor$ in \eqref{eq:erdos-turan-discrepancy} and using
Lemma~\ref{lem:uniform-quadratic-weyl-sums}, we obtain
\begin{equation}\label{eq:phase-factor-discrepancy}
\mathfrak D(\mu_{D,N})\leq
D_B(N):=N^{-1/16}+2W(\gamma)(1+\log N)N^{-1/8}.
\end{equation}
To turn this root count into a product bound, we use the
bounded-variation estimate
\begin{equation}\label{eq:bounded-variation-discrepancy}
\left|\int_{\T}f\,d\mu-\int_{\T}f(t)\,dt\right|
\leq\mathfrak D(\mu)\Var_{\T}(f),
\end{equation}
valid for $f\in BV(\T)\cap C(\T)$, which can be found in
Appendix~\ref{app:bounded-variation-discrepancy}. The following lemma applies it to a truncated
logarithm. We state both bounds here because the lower bound will also
be needed for $Q_N$.

\begin{lem}\label{lem:circle-product-bounds}
Let $\xi_1,\ldots,\xi_N\in\T$, and define their normalized counting measure
\begin{align*}
\mu_N(\boA)=\frac1N\sum_{j=1}^N \bm 1_{\boA}(\xi_j)\ \text{for $\boA\subset\T$}.
\end{align*}
Suppose that $\mu_N$ has discrepancy $\mathfrak D(\mu_N)\leq D$, where $0<D\leq1/2$. For $|w|=1$, set
\begin{align*}
Q(w)=\prod_{j=1}^N\left(1-we^{-2\pi\im\xi_j}\right).
\end{align*}
Then
\begin{equation}\label{eq:circle-product-upper-bound}
\log|Q(w)|\leq4ND\log(2/D),
\end{equation}
If $\left|w-e^{2\pi\im\xi_j}\right|\geq\varepsilon$ for every $j$, where $0<\varepsilon\leq1$, then
\begin{equation}\label{eq:circle-product-lower-bound}
\log|Q(w)|\geq-4ND\big[\log(2/D)+\log(1/\varepsilon)\big].
\end{equation}
\end{lem}

\begin{proof}
Let $w=e^{2\pi\im\theta}$ and define
\begin{align*}
	\widetilde\mu_N(\boA)=\mu_N(\boA+\theta),
	\ \boA\subset\T,
\end{align*} 
then $	\mathfrak D(\widetilde\mu_N)=\mathfrak D(\mu_N)$. For
\begin{align*}
	\ell(t)=\log|1-e^{2\pi\im t}|
	=\log\big(2|\sin\pi t|\big),
\end{align*}
we have
\begin{align*}
	\frac1N\log|Q(w)|
	=\int_{\T}\ell\,d\widetilde\mu_N,
	\ 
	\int_{\T}\ell(t)\,dt=0.
\end{align*}
To apply the discrepancy estimate, we truncate the logarithmic
singularity by setting
\begin{align*}
	\ell_h(t)=\max\{\ell(t),\log h\},
	\ 0<h\leq1.
\end{align*}
By Appendix~\ref{app:logarithmic-kernel}, we have
\begin{align*}
	0\leq\int_{\T}\ell_h(t)\,dt\leq h,\ \Var_{\T}(\ell_h)=2\log(2/h).
\end{align*}
Thus by \eqref{eq:bounded-variation-discrepancy}, we get
\begin{equation}\label{eq:truncated-log-discrepancy}
	\left|\int_{\T}\ell_h\,d\widetilde\mu_N
	-\int_{\T}\ell_h(t)\,dt\right|
	\leq2D\log(2/h).
\end{equation}
Taking $h=D$ and using $\ell\leq\ell_h$, we obtain
\begin{align*}
	\frac1N\log|Q(w)|\leq D+2D\log(2/D)\leq4D\log(2/D).
\end{align*}
If $|w-e^{2\pi\im\xi_j}|\geq\varepsilon$ for every $j$, take
$h=\min\{D,\varepsilon\}\geq D\ep$. Then we have
\begin{align*}
	\ell(\xi_j-\theta)=\log|1-e^{2\pi\im(\xi_j-\theta)}|=\log|w-e^{2\pi\im\xi_j}|\geq\log\varepsilon\ge\log h,
\end{align*}
which implies $\ell_h(\xi_j-\theta)=\ell(\xi_j-\theta)$ for all $j$. Hence, one has
\begin{align*}
	\frac1N\log|Q(w)|\geq-2D\log(2/h)\geq-2D\big[\log(2/D)+\log(1/\varepsilon)\big].
\end{align*}
This proves \eqref{eq:circle-product-upper-bound} and \eqref{eq:circle-product-lower-bound}. We complete the proof.
\end{proof}

Applying \eqref{eq:circle-product-upper-bound} to the roots of $B_N$ gives
\begin{align*}
\log|B_N(w)|\leq4ND_B(N)\log\frac{2}{D_B(N)},
\ D_B(N)\leq\tfrac12,
\end{align*}
We will put it in a common form with the bound for $Q_N$ below.

For the rest of the proof, fix
\begin{equation}\label{eq:discrepancy-and-resonance-exponents}
\sigma=\frac1{32},\  a=\frac1{128},\
\nu=1-\sigma+a=\frac{125}{128}<1.
\end{equation}
With discrepancy $N^{-\sigma}$ and root separation
$ce^{-N^a}$, the logarithmic error is $O(N^\nu)$, where $\nu=1-\sigma+a<1$.

\subsection[The free Cayley factor in Q]{The free Cayley factor in \texorpdfstring{$Q_N$}{Q}}\label{subsec:free-cayley-kernel}

For $Q_N$, the roots are determined by the eigenvalues
$e^{2\pi\im\xi_j}$ of $\sU_{E,N}$. Put
\begin{align*}
\mu_{E,N}(\boA)=\frac1N\sum_{j=1}^N \bm 1_{\boA}(\xi_j)\ \text{for $\boA\subset\T$},\
\hat\mu_{E,N}(k)=\frac1N\Tr(\sU_{E,N}^k).
\end{align*}
Again the root angles are $\tfrac12-\xi_j$. To use \eqref{eq:erdos-turan-discrepancy},
we must therefore estimate these traces. We first replace the free
Cayley factor by a convolution with short jumps, controlling the
boundary and truncation errors.

Define
\begin{equation}\label{eq:infinite-volume-cayley-transform}
\sR_E=\big(\lambda\Delta+(\im-E)\sI\big)^{-1},
\ \sC_E=\sI-2\im\sR_E.
\end{equation}
By the spectral theorem, $\|\sR_E\|\leq1$ and $\sC_E$ is unitary.

\begin{lem}\label{lem:free-resolvent-decay}
Let
\begin{equation}\label{eq:free-resolvent-decay-rate}
\mu_\lambda=\log\left(1+\frac1{4\lambda}\right)>0.
\end{equation}
Then we get
\begin{equation}\label{eq:free-resolvent-kernel-bound}
|\sR_E(n,m)|\leq2e^{-\mu_\lambda|n-m|}.
\end{equation}
Consequently, $\sC_E(n,m)=c_E(n-m)$, with
$|c_E(r)|\leq4e^{-\mu_\lambda|r|}$ for $r\neq0$.
\end{lem}

The decay estimate follows from an exponential-weight
conjugation of the resolvent, which can be found in Appendix~\ref{app:determinant-factor-estimates}.
We next compare $\sC_E$ with its finite-volume counterpart.

\begin{lem}\label{lem:cayley-boundary-comparison}
	Regard $\sP_N\sC_E\sP_N$ as acting on
	$\ell^2(\lg_N)$. Then
	\begin{equation}\label{eq:cayley-compression-trace-norm}
		\|\sC_{E,N}-\sP_N\sC_E\sP_N\|_{\mathfrak S_1}\leq8.
	\end{equation}
	Consequently, for every $k\geq1$,
	\begin{equation}\label{eq:compressed-cayley-trace-error}
		\left|\Tr(\sU_{E,N}^k)
		-\Tr\big((\sP_N\sC_E\sP_N\sD_N)^k\big)\right|
		\leq8k.
	\end{equation}
\end{lem}

With $\sP=\sP_N$, $\sQ=\sI-\sP$, and
$\sR_{E,N}=(\sK_{E,N}+\im\sI_N)^{-1}$, we have
\begin{equation}\label{eq:resolvent-compression-identity}
	\sR_{E,N}-\sP\sR_E\sP
	=\sR_{E,N}\sP\lambda\Delta\sQ\sR_E\sP.
\end{equation}
Since $\sP\Delta\sQ$ has rank at most two, the
resolvent difference has rank at most two. The estimates
\eqref{eq:cayley-compression-trace-norm}--\eqref{eq:compressed-cayley-trace-error} follow from the trace-norm
bound and the telescoping identity, respectively, which can be found in Appendix~\ref{app:resolvent-compression-and-trace-norms}.

Let $\sC_E^{[M]}$ be the convolution operator with kernel
$c_E(r)\mathbf1_{\{|r|\leq M\}}$. By Lemma~\ref{lem:free-resolvent-decay}, we get
\begin{align*}
	\|\sC_E-\sC_E^{[M]}\|\leq\sum_{|r|>M}|c_E(r)|\leq\frac{8e^{-\mu_\lambda M}}
	{e^{\mu_\lambda}-1}=32\lambda e^{-\mu_\lambda M}.
\end{align*}
Choose
\begin{align*}
	M_N&=\left\lceil
	\frac{12\log(N+2)+\log(32\lambda+1)}
	{\mu_\lambda}\right\rceil,\\
	A(\lambda)&=\frac{12}{\mu_\lambda}
	+\frac{\log(32\lambda+1)}{\mu_\lambda\log3}
	+\frac2{\log3}.
\end{align*}
Then, we obtain
\begin{equation}\label{eq:cayley-truncation-bounds}
	\begin{aligned}
		M_N+1&\leq A(\lambda)\log(N+2),\\
		\|\sC_E-\sC_E^{[M_N]}\|
		&\leq\varepsilon_N:=N^{-12},\\
		\|\sC_E^{[M_N]}\|&\leq1+\varepsilon_N.
	\end{aligned}
\end{equation}

\subsection[Traces and the roots of Q]{Traces and the roots of \texorpdfstring{$Q_N$}{Q}}\label{subsec:cayley-traces-and-root-discrepancy}

The trace of $\sU_{E,N}^k$ is a sum over closed paths.
The quadratic phase along each path separates into a
common term depending on the starting point and a
path-dependent remainder.

Fix $M=M_N$ and define the diagonal operator
$\sD_{\alpha,\beta}$ by
\begin{align*}
	\sD_{\alpha,\beta}
	=e^{2\pi\im(\alpha n^2+\beta n)}\delta_{n,n'},
	\ n,n'\in\Z.
\end{align*}

\begin{lem}\label{lem:closed-path-phase-decomposition}
For each $k\geq1$, there is a trigonometric polynomial $f_k$ of degree
at most $k^2M$ such that, for every $n\in\Z$,
\begin{equation}\label{eq:closed-path-diagonal-entry}
\big[(\sC_E^{[M]}\sD_{\alpha,\beta})^k\big](n,n)
=e^{2\pi\im(k\alpha n^2+k\beta n)}f_k(\beta+2\alpha n).
\end{equation}
If $N\geq2$ and $k\leq N^{1/16}$, then
\begin{equation}\label{eq:closed-path-fourier-norm}
\sum_\ell|\hat f_k(\ell)|\leq4k\sqrt{M+1}.
\end{equation}
\end{lem}

\begin{proof}
	Write the visited points as $n+b_0,\ldots,n+b_k$, where
	$b_0=b_k=0$ and $|b_j-b_{j-1}|\leq M$. Then
	\begin{align*}
		\alpha(n+b_j)^2+\beta(n+b_j)
		=\alpha n^2+\beta n
		+(\beta+2\alpha n)b_j+\alpha b_j^2.
	\end{align*}
	Consequently, \eqref{eq:closed-path-diagonal-entry} holds with
	\begin{equation}\label{eq:closed-path-polynomial}
		f_k(t)=
		\sum_{\substack{b_0=b_k=0\\|b_j-b_{j-1}|\leq M}}
		\left[\prod_{j=1}^kc_E(b_{j-1}-b_j)\right]
		e^{2\pi\im\left(
		t\sum_{j=1}^kb_j+\alpha\sum_{j=1}^kb_j^2
		\right)}.
	\end{equation}
	Since $|b_j|\leq jM$, we have
	\begin{align*}
	\left|\sum_{j=1}^kb_j\right|\leq k^2M.
	\end{align*}
	Thus $f_k$ has at most $2k^2M+1$ nonzero Fourier coefficients.
	
	Moreover, we get
	\begin{align*}
		f_k(t)=
		\langle\delta_0,
		(\sC_E^{[M]}\sD_{\alpha,t})^k\delta_0\rangle.
	\end{align*}
	By \eqref{eq:cayley-truncation-bounds}, we obtain
	\begin{align*}
			\|f_k\|_\infty
		\leq(1+\varepsilon_N)^k
		\leq e^{k\varepsilon_N}\leq2.
	\end{align*}
From the Parseval's identity, we have
\begin{align*}
	\sum_{\ell\in\mathbb Z}
	|\widehat f_k(\ell)|^2=\|f_k\|_{L^2(\T)}^2\le 4.
\end{align*}
Hence, by Cauchy-Schwarz inequality, we obtain
\begin{align*}
	\sum_{\ell\in\mathbb Z}|\widehat f_k(\ell)|
	&\leq\sqrt{2k^2M+1}
	\left(\sum_{\ell\in\mathbb Z}
	|\widehat f_k(\ell)|^2\right)^{1/2}\\
	&\leq2\sqrt{2k^2M+1}
	\leq4k\sqrt{M+1}.
\end{align*}
This proves \eqref{eq:closed-path-fourier-norm}. We have finished the proof.
\end{proof}

Each Fourier coefficient of $f_k$ produces a quadratic Weyl sum with the same quadratic coefficient $k\alpha$ and a modified linear coefficient.

\begin{prop}\label{prop:cayley-trace-estimate}
If $\alpha\in\boD_\gamma$, then, for every $N\geq2$ and
$1\leq k\leq N^{1/16}$,
\begin{equation}\label{eq:normalized-cayley-trace-bound}
\begin{aligned}
\frac{|\Tr(\sU_{E,N}^k)|}{N}
\leq\ &4W(\gamma)\sqrt{A(\lambda)}\,
 k\sqrt{\log(N+2)}\,N^{-1/8}\\
&\ +8A(\lambda)\frac{k\log(N+2)}N+2kN^{-12}.
\end{aligned}
\end{equation}
\end{prop}

\begin{proof}
	By Lemma~\ref{lem:cayley-boundary-comparison} and the truncation estimate
	\eqref{eq:cayley-truncation-bounds}, together with the boundary path comparison
	in Appendix~\ref{app:finite-trace-path-comparison}, we have
	\begin{equation}\label{eq:finite-trace-path-comparison}
		\begin{aligned}
			\left|\Tr(\sU_{E,N}^k)-
			\sum_{n=0}^{N-1}
			e^{2\pi\im(k\alpha n^2+k\beta n)}
			f_k(\beta+2\alpha n)\right|\leq8k+2Nk\varepsilon_N+8kM.
		\end{aligned}
	\end{equation}
	Expanding $f_k$ in Fourier series, the normalized path sum becomes
	\begin{align*}
		\sum_{\ell\in\mathbb Z}\widehat f_k(\ell)
		e^{2\pi\im\ell\beta}
		\frac1N\sum_{n=0}^{N-1}
		e^{2\pi\im(k\alpha n^2+(k\beta+2\ell\alpha)n)}.
	\end{align*}
	By Lemmas~\ref{lem:uniform-quadratic-weyl-sums} and~\ref{lem:closed-path-phase-decomposition}, its absolute value
	is bounded by
	\begin{align*}
			4W(\gamma)k\sqrt{M+1}\,N^{-1/8}.
	\end{align*}
	Combining this with \eqref{eq:finite-trace-path-comparison} and using
	\begin{align*}
			\varepsilon_N=N^{-12},\ 	M+1\leq A(\lambda)\log(N+2),
	\end{align*}
	gives \eqref{eq:normalized-cayley-trace-bound}. This completes the proof.
\end{proof}

We can now apply \eqref{eq:erdos-turan-discrepancy} to $\mu_{E,N}$. The following common threshold also includes the
quadratic phase measure considered above.

\begin{cor}\label{cor:determinant-root-discrepancy}
	Fix $\lambda>0$ and $0<\gamma<1/2$. Set
	\begin{align*}
		C_{\mathrm{disc}}(\lambda,\gamma)
		&=5+8W(\gamma)\sqrt{A(\lambda)}+16A(\lambda),\\
		N_0(\lambda,\gamma)
		&=\max\left\{2^{32},\left\lceil
		\big(\sqrt{32}\,C_{\mathrm{disc}}(\lambda,\gamma)\big)^{64}
		\right\rceil\right\}.
	\end{align*}
	If $\alpha\in\boD_\gamma$ and $N\geq N_0(\lambda,\gamma)$, then
	\begin{equation}\label{eq:uniform-root-discrepancy}
		\mathfrak D(\mu_{E,N}),\
		\mathfrak D(\mu_{D,N})
		\leq N^{-\sigma}.
	\end{equation}
	The root counting measures of $Q_N$ and $B_N$ satisfy
	the same discrepancy bound.
\end{cor}

\begin{proof}
	Taking $K=\lfloor N^{1/16}\rfloor$ in \eqref{eq:erdos-turan-discrepancy}
	and using Proposition~\ref{prop:cayley-trace-estimate}, we obtain
	\begin{align*}
		\mathfrak D(\mu_{E,N})
		&\leq N^{-1/16}
		+8W(\gamma)\sqrt{M+1}\,N^{-1/16}\\
		&\ +16(M+1)N^{-15/16}+4N^{-191/16}\\
		&\leq C_{\mathrm{disc}}(\lambda,\gamma)
		N^{-1/16}\sqrt{\log(N+2)},
	\end{align*}
	where we used \eqref{eq:cayley-truncation-bounds} and
	$1\leq\log(N+2)\leq N$.
	By \eqref{eq:quadratic-weyl-sum-bound}, the Fourier coefficients of
	$\mu_{D,N}$ satisfy
	\begin{align*}
			|\widehat\mu_{D,N}(k)|
		\leq W(\gamma)N^{-1/8},
		\ 1\leq k\leq K.
	\end{align*}
	The same application of \eqref{eq:erdos-turan-discrepancy} therefore gives
	the same discrepancy bound for $\mu_{D,N}$.
	
	Since $\log(N+2)\leq32N^{1/32}$,
	we conclude that
	\begin{align*}
		\mathfrak D(\mu_{E,N}),\,
		\mathfrak D(\mu_{D,N})\leq\sqrt{32}\,C_{\mathrm{disc}}(\lambda,\gamma)
		N^{-3/64}\leq N^{-1/32},
	\end{align*}
	for $N\geq N_0(\lambda,\gamma)$. The root counting measures of $Q_N$ and $B_N$
	inherit the same bounds under
	$\xi\mapsto\frac12-\xi$. We have concluded the proof. 
\end{proof}

\subsection[Upper and lower bounds for Q]{Upper and lower bounds for \texorpdfstring{$Q_N$}{Q}}

We now apply Lemma~\ref{lem:circle-product-bounds} to $Q_N$ and $B_N$,
including intervals of arbitrary length.

\begin{cor}	\label{cor:phase-and-spectral-factor-upper-bounds}
	Fix $\lambda>0$ and $0<\gamma<1/2$, and suppose that
	$\alpha\in\boD_\gamma$. Let
	$N_0=N_0(\lambda,\gamma)$ be as in
	Corollary~\ref{cor:determinant-root-discrepancy}, and define
	\begin{align*}
		F_{N_0}(n)
		=N_0\log2+4n^{1-\sigma}\log(n+2),
		\ n\geq0.
	\end{align*}
	Then, for every integer $n\geq0$, $E\in\R$,
	$|w|=1$, and $\beta\in\T$,
	\begin{equation}\label{eq:phase-and-spectral-log-upper-bound}
		\log|Q_n(w;E)|,\ \log|B_n(w)|
		\leq F_{N_0}(n),
	\end{equation}
	where $Q_0=B_0=1$.
\end{cor}

\begin{proof}
	For $n\geq N_0$, Lemma~\ref{lem:circle-product-bounds} with
	$D=n^{-\sigma}$ gives
	\begin{align*}
		\log|Q_n(w;E)|,\ \log|B_n(w)|\leq4n^{1-\sigma}\log(2n^\sigma)\leq4n^{1-\sigma}\log(n+2).
	\end{align*}
	For $1\leq n<N_0$, the bound follows from
	$|1+wu|\leq2$ for $|w|=|u|=1$.
	The case $n=0$ is immediate. This concludes the proof.
\end{proof}
For a lower bound on $Q_N$, we need to separate $w$
from its roots. We use the cleared matrix in \eqref{eq:denominator-cleared-factorization}. Set
\begin{align*}
	\sL_E(z)=(\sK_{E,N}+\im\sI_N)+(\sK_{E,N}-\im\sI_N)z\sD_{N}.
\end{align*}
Since
\begin{align*}
	\det\sL_E(z)=A_N(E)Q_N(z;E),
\end{align*}
every root $w_0$ of $Q_N(\cdot;E)$ admits a unit vector
$v\in\ker\sL_E(w_0)$. Setting
$u_0=(\sI_N+w_0\sD_{N})v$, we obtain
\begin{equation}\label{eq:cleared-matrix-kernel-identity}
	(\sK_{E,N}-\im\sI_N)u_0=-2\im v.
\end{equation}
In particular, $u_0\neq0$, even when $w_0$ is a tangent pole. At the current pole-free phase $w$, let
$u=(\sI_N+w\sD_{N})v$. Subtracting $\sL_E(w_0)v=0$ from \eqref{eq:denominator-cleared-factorization}
applied to $v$ at $w$, we obtain
\begin{align*}
	(\sH_N-E\sI_N)u
	=(\sK_{E,N}-\im\sI_N)(w-w_0)\sD_Nv.
\end{align*}
Moreover, since $\sD_N$ is unitary and $\|v\|=1$, we have
\begin{align*}
	\|u-u_0\|=|w-w_0|.
\end{align*}
The residual is proportional to $|w-w_0|$, while
\eqref{eq:cleared-matrix-kernel-identity} controls $\|u_0\|$ from below.

\begin{lem}\label{lem:spectral-gap-to-phase-root-separation}
	Suppose that $|E|\leq E_*$ with $E_*\geq0$, and let
	$C=1+E_*+2\lambda$.
	If $w=e^{2\pi\im\theta}$ is pole-free and
	\begin{align*}
		\dist\big(E,\sigma(\sH_N)\big)\geq\delta,
		\ 0<\delta\leq1,
	\end{align*}
	then every root $w_0$ of $Q_N(\cdot;E)$ satisfies
	\begin{equation}\label{eq:phase-root-separation-bound}
		|w-w_0|\geq C^{-2}\delta.
	\end{equation}
	This includes roots that are also zeros of $B_N$.
\end{lem}

The proof is given in Appendix~\ref{app:phase-root-separation-proof}.
In particular, the bound is independent of the size of
the tangent potential at the current phase.

If $N\geq N_0$, $\alpha\in\boD_\gamma$, $|E|\leq E_*$
and
\begin{align*}
	\dist(E,\sigma(\sH_N))\geq e^{-N^a},
\end{align*}
then Lemma~\ref{lem:spectral-gap-to-phase-root-separation} gives
\begin{align*}
	|w-w_0|\geq C^{-2}e^{-N^a}
\end{align*}
for every root $w_0$ of $Q_N(\cdot;E)$.
Applying \eqref{eq:circle-product-lower-bound} with $D=N^{-\sigma}$ and
$\varepsilon=C^{-2}e^{-N^a}$, we obtain
\begin{align}\label{eq:nonresonant-spectral-factor-lower-bound}
	\nonumber	\log|Q_N(w;E)|
		&\geq-4N^{1-\sigma}
		\big[N^a+\log(N+2)+2\log C\big]\\
		&\geq-(524+8\log C)N^\nu.
\end{align}
Here we used $\log(N+2)<130N^a$.

\section{Green function decay}\label{sec:green-function-decay}

The cofactor formula combines the three determinant factors.
The $B$ factors require only upper bounds, whereas $Q$
requires both upper and lower bounds.

Let $\boI=[b,c]\cap\Z$ and $x\leq y$ in $\boI$. Write $d=y-x$,
$\boI_-=[b,x-1]\cap\Z$, and $\boI_+=[y+1,c]\cap\Z$.
Direct computation shows
\begin{equation}\label{eq:green-function-cofactor-formula}
\sG_{\boI}(x,y;E)=(-\lambda)^d
\frac{\det(\sH_{\boI_-}-E\sI_{\boI_-})\det(\sH_{\boI_+}-E\sI_{\boI_+})}{\det(\sH_{\boI}-E\sI_{\boI})}.
\end{equation}

Applying \eqref{eq:three-factor-determinant} to the three determinants with the
same phase $w$, the $\sB$ factors cancel outside $[x,y]$,
leaving
\begin{equation}\label{eq:tangent-denominator-cancellation}
	\frac{B_{\boI}(w)}
	{B_{\boI_-}(w)B_{\boI_+}(w)}
	=\prod_{j=x}^y
	\big(1+e^{2\pi\im(\theta+\alpha j^2)}\big).
\end{equation}
The remaining $Q$ factors are
$Q_{\boI_-}$ and $Q_{\boI_+}$ in the numerator,
and $Q_{\boI}$ in the denominator.

\begin{prop}\label{prop:nonresonant-green-function-decay}
Fix $\lambda>0$, $\gamma>0$, and $0\leq E_*<\infty$. Take $N_0$ from
Corollary~\ref{cor:determinant-root-discrepancy} and $C$ from Lemma~\ref{lem:spectral-gap-to-phase-root-separation}. Define
\begin{align*}
M(\lambda,\gamma,E_*)
 &=2084+8\log C+3J(\lambda)+3N_0\log2,\\
N_1(\lambda,\gamma,E_*)
 &=\max\left\{N_0,\left\lceil
 \left(\frac{20M(\lambda,\gamma,E_*)}{g_\lambda}\right)^{1/(1-\nu)}
 \right\rceil\right\}.
\end{align*}
If $\alpha\in\boD_\gamma$ and the potential is finite
on a finite interval $\boI\subset\Z$ of length $N\geq N_1$
satisfies
\begin{equation}\label{eq:finite-volume-nonresonance}
\dist\big(E,\sigma(\sH_{\boI})\big)\geq e^{-N^a},
\ |E|\leq E_*,
\end{equation}
then
\begin{equation}\label{eq:off-diagonal-green-function-bound}
|\sG_{\boI}(x,y;E)|\leq e^{-\tfrac12g_\lambda|x-y|},\ |x-y|\geq N/10.
\end{equation}
\end{prop}

\begin{proof}
	By \eqref{eq:translated-phase-parameters}, the local estimates apply to $\boI$. By symmetry, assume $x\leq y$,
	and put $d=y-x$ and $n_\pm=\#\boI_\pm$.
	If the numerator in \eqref{eq:green-function-cofactor-formula} vanishes, the claim is
	immediate. Otherwise, apply the Cayley factorization to
	\eqref{eq:green-function-cofactor-formula}.
	
	For the free factors, write
	\begin{align*}
		r_n=\log|A_n(E)|
		-n\big(\log\lambda+g(E,\lambda)\big),
		\ |r_n|\leq J(\lambda).
	\end{align*}
	Since $n_-+n_+=N-d-1$, Lemma~\ref{lem:free-determinant-growth} gives
	\begin{align*}
		&d\log\lambda+\log|A_{n_-}(E)|
		+\log|A_{n_+}(E)|-\log|A_N(E)|\\
		&=-\log\lambda-(d+1)g(E,\lambda)
		+r_{n_-}+r_{n_+}-r_N\\
		&\leq-g_\lambda d+3J(\lambda),
	\end{align*}
	since $|r_n|\leq J(\lambda)$ and
	$\lambda e^{g_\lambda}=\frac{1+\sqrt{1+4\lambda^2}}{2}\geq1$.
	For the $B$ factors, \eqref{eq:tangent-denominator-cancellation} and
	Corollary~\ref{cor:phase-and-spectral-factor-upper-bounds} yield
	\begin{align*}
		\log\left|
		\frac{B_{\boI}(w)}
		{B_{\boI_-}(w)B_{\boI_+}(w)}
		\right|
		\leq N_0\log2+4N^{1-\sigma}\log(N+2).
	\end{align*}
		For the $Q$ factors, Corollary~\ref{cor:phase-and-spectral-factor-upper-bounds} gives
	\begin{align*}
		&\log|Q_{\boI_-}(w;E)|
		+\log|Q_{\boI_+}(w;E)|\\
		&\leq2N_0\log2+8N^{1-\sigma}\log(N+2).
	\end{align*}
The denominator is bounded from below by \eqref{eq:nonresonant-spectral-factor-lower-bound}.
Consequently, the total contribution of the three
factors is at most
\begin{align*}
	&3N_0\log2+12N^{1-\sigma}\log(N+2)
	+(524+8\log C)N^\nu\\
	&\leq
	\big(2084+8\log C+3N_0\log2\big)N^\nu,
\end{align*}
where we used $\log(N+2)<130N^a$ and
$\nu=1-\sigma+a$.

Combining these estimates with the free determinant bound gives
\begin{equation}\label{eq:green-function-log-bound}
	\log|\sG_{\boI}(x,y;E)|
	\leq-g_\lambda d+M(\lambda,\gamma,E_*)N^\nu.
\end{equation}
By the choice of $N_1$, we have
\begin{align*}
	M(\lambda,\gamma,E_*)N^\nu
	\leq \frac{g_\lambda N}{20}
	\leq \frac{g_\lambda d}{2}.
\end{align*}
Hence \eqref{eq:green-function-log-bound} yields \eqref{eq:off-diagonal-green-function-bound}. The proof is now complete.
\end{proof}

\section{From spectral separation to localization}\label{sec:spectral-separation-and-localization}

\subsection{Excluding double resonances}

Since a nonzero solution is uniquely determined by its
values at two consecutive sites, at least one of $\psi(0)$ and $\psi(1)$ must be nonzero. We choose
$c\in\{0,1\}$ so that $\psi(c)\neq0$.
For each sufficiently distant $R$, set
$N=\lfloor |R|/4\rfloor\geq2$. Define
\begin{align}
	\boA_N&=\{\pm(4N+j):j=0,1,2,3\},\label{eq:distant-block-centers}\\
	\boI_{c,N}
	&=[c-N,c+N]\cap\Z,\ c\in\{0,1\},\nonumber\\
	\mathcal J_{R,N}
	&=[R-N,R+N]\cap\Z,\ R\in\boA_N.
	\label{eq:distant-resonance-block}
\end{align}

If
\begin{align*}
	\dist(E,\sigma(\sH_{\boI_{c,N}}))<\tau,
	\
	\dist(E,\sigma(\sH_{\mathcal{J}_{R,N}}))<\tau,
\end{align*}
then
\begin{align*}
	\dist\big(
	\sigma(\sH_{\boI_{c,N}}),
	\sigma(\sH_{\mathcal{J}_{R,N}})
	\big)<2\tau.
\end{align*}
Thus a separation estimate for the two finite-volume
spectra excludes simultaneous resonances.

\begin{lem}\label{lem:two-block-spectral-separation-measure}
For $\lambda>0$, $N\geq2$, $c\in\{0,1\}$, $R\in\boA_N$, and
$\delta>0$,
\begin{equation}\label{eq:two-block-spectral-gap-measure}
\mes\left\{(\theta,\alpha)\notin\boP:
\dist\big(\sigma(\sH_{\boI_{c,N}}),\sigma(\sH_{\mathcal{J}_{R,N}})\big)<\delta\right\}
\leq200N^3\delta.
\end{equation}
The bound is uniform in $\lambda$ and requires no arithmetic
condition on $\alpha$.
\end{lem}

\begin{proof}
The potential increases with $\theta$ on $\boI_{c,N}$ and
$\mathcal J_{R,N}$. To obtain opposite eigenvalue motion,
we vary $\theta$ and $\alpha$ simultaneously. Since
$n^2\leq(N+1)^2$ for $n\in\boI_{c,N}$ and
$n^2\geq9N^2$ for $n\in\mathcal J_{R,N}$, choose $4N^2$
between these bounds and set
\begin{equation}\label{eq:parameter-shear}
	\theta=\vartheta-4N^2t\pmod1,\ \alpha=t.
\end{equation}
The phase at point $n\in\Z$ becomes
$\vartheta+s_nt$, where $s_n=n^2-4N^2$. Thus, we get
\begin{align}
	s_n&\leq-N^2,\ n\in\boI_{c,N},\label{eq:central-block-shear-slopes}\\
	s_n&\geq5N^2,\ n\in\mathcal J_{R,N}.\label{eq:distant-block-shear-slopes}
\end{align}

On each pole-free interval in $t$, let
$\eta_j(t)$ and $\zeta_k(t)$ denote the analytic eigenvalue
branches of $\sH_{\boI_{c,N}}$ and
$\sH_{\mathcal J_{R,N}}$, respectively.
By Rellich's theorem and the Feynman-Hellmann formula,
\begin{align*}
	\eta_j'(t)
	&=\pi\sum_{n\in\boI_{c,N}}
	|\psi_j(n)|^2s_n
	\sec^2\pi(\vartheta+s_nt),\\
	\zeta_k'(t)
	&=\pi\sum_{n\in\mathcal J_{R,N}}
	|\phi_k(n)|^2s_n
	\sec^2\pi(\vartheta+s_nt),
\end{align*}
where $\psi_j$ and $\phi_k$ are normalized eigenvectors.
Consequently, we obtain
\begin{align*}
	\eta_j'(t)\leq-\pi N^2,\ \zeta_k'(t)\geq5\pi N^2,
\end{align*}
and hence
\begin{align*}
		(\zeta_k-\eta_j)'(t)\geq6\pi N^2.
\end{align*}
Hence, by the mean value theorem,
\begin{align*}
	\mes\{t:|\zeta_k(t)-\eta_j(t)|<\delta\}
	\leq\frac{\delta}{3\pi N^2}.
\end{align*}

Since $s_n\in\Z\setminus\{0\}$, the map
$t\mapsto s_nt\pmod1$ is an $|s_n|$-to-one
covering of $\T$. Hence each point $n$ contributes
exactly $|s_n|$ poles.
Each block has length $L=2N+1$. Summing over the $L^2$ eigenvalue
pairs and all pole-free intervals, and integrating in $\vartheta$, gives
\eqref{eq:two-block-spectral-gap-measure}, which can be found in Appendix~\ref{app:shear-pole-count}. We close the proof here.
\end{proof}

Set
\begin{equation}\label{eq:resonance-threshold}
	\tau_N=e^{-(2N+1)^a}.
\end{equation}
Let $\mathcal{R}_N^\lambda$ be the union of the events
\begin{align*}
	\bigcup_{\substack{c\in\{0,1\}\\R\in\boA_N}}
	\left\{(\theta,\alpha)\notin\boP:
	\dist\big(\sigma(\sH_{\boI_{c,N}}),
	\sigma(\sH_{\mathcal J_{R,N}})\big)<2\tau_N\right\}.
\end{align*}
Since there are $16$ pairs
$(c,R)$, Lemma~\ref{lem:two-block-spectral-separation-measure} gives
\begin{equation}\label{eq:double-resonance-summability}
	\mes(\mathcal{R}_N^\lambda)
	\leq6400N^3e^{-(2N+1)^a},
	\ 
	\sum_{N\geq2}\mes(\mathcal{R}_N^\lambda)<\infty.
\end{equation}
\begin{prop}[An energy-independent set of good parameters]
	\label{prop:energy-independent-good-parameters}
	Let
	\begin{equation}\label{eq:good-parameter-set}
		\boG_\lambda=
		\left[\T\times\bigcup_{j\geq3}\boD_{1/j}\right]
		\setminus\left(\boP\cup
		\limsup_{N\to\infty}\mathcal R_N^\lambda\right).
	\end{equation}
	Then $\mes\boG_\lambda=1$. Moreover, for every
	$(\theta,\alpha)\in\boG_\lambda$, there exists $N_2$
	such that, for all $N\geq N_2$, $c\in\{0,1\}$, and
	$R\in\boA_N$,
	\begin{equation}\label{eq:eventual-two-block-spectral-separation}
		\dist\big(
		\sigma(\sH_{\boI_{c,N}}),
		\sigma(\sH_{\mathcal J_{R,N}})
		\big)\geq2\tau_N.
	\end{equation}
\end{prop}

\begin{proof}
	By \eqref{eq:diophantine-exceptional-measure}, we have
	\begin{align*}
		\mes\left(\bigcup_{j\geq3}\boD_{1/j}\right)=\lim\limits_{j\to\infty}\mes(\boD_{1/j})=1.
	\end{align*}
	By \eqref{eq:double-resonance-summability} and the Borel-Cantelli Lemma,
	\begin{align*}
		\mes(\limsup_{N\to\infty}\mathcal R_N^\lambda)=0.
	\end{align*}
	Since $\mes(\boP)=0$, the first assertion follows.
	The second follows directly from the definition of
	$\boG_\lambda$. This concludes the proof.
\end{proof}

 In particular, if
\begin{align*}
	\dist(E,\sigma(\sH_{\boI_{c,N}}))<\tau_N,
\end{align*}
then
\begin{align*}
	\dist(E,\sigma(\sH_{\mathcal J_{R,N}}))>\tau_N,\ R\in\boA_N.
\end{align*}

We next show that every nonzero polynomially bounded
solution produces such a central resonance.

\subsection{Proof of Theorem \ref{thm:all-coupling-localization}}
\label{subsec:proof-of-localization}
As an application of Green’s function estimates of the previous section, we prove Theorem \ref{thm:all-coupling-localization}  below.

\begin{proof}[Proof of Theorem \ref{thm:all-coupling-localization}]
	Fix $\lambda>0$ and $(\theta,\alpha)\in\boG_\lambda$.
	Let $E\in\R$ and let $\psi\neq0$ be a polynomially bounded
	solution of
	\begin{equation}\label{eq:generalized-eigenfunction-equation}
		\sH_{\lambda,\theta,\alpha}\psi=E\psi,
	\end{equation}
	with
	\begin{align*}
		|\psi(n)|\leq (1+|n|)^2.
	\end{align*}
	
	Choose $\gamma>0$ such that $\alpha\in\boD_\gamma$,
	and fix $E_*\geq|E|$. Since $\lambda>0$, the recurrence
	determines $\psi$ from its values at two consecutive points.
	Thus, there exists $c\in\{0,1\}$ such that $\psi(c)\neq0$.
	
	For all sufficiently large $N$, Propositions~\ref{prop:nonresonant-green-function-decay}
	and~\ref{prop:energy-independent-good-parameters} apply. We first claim that
	\begin{equation}\label{eq:central-block-resonance}
		\dist(E,\sigma(\sH_{\boI_{c,N}}))<\tau_N.
	\end{equation}
	Indeed, otherwise the Green's function estimate of
	Proposition~\ref{prop:nonresonant-green-function-decay} and the Poisson formula give
	\begin{align*}
		|\psi(c)|
		&\leq\lambda e^{-g_\lambda N/2}
		\bigl(|\psi(c-N-1)|+|\psi(c+N+1)|\bigr)\\
		&\leq2\lambda (N+3)^2 e^{-g_\lambda N/2},
	\end{align*}
	which tends to zero as $N\to\infty$, contradicting
	$\psi(c)\neq0$.
	
	Now let $R\in\Z$ be sufficiently large and set
	$N=\lfloor |R|/4\rfloor$, so that $R\in\boA_N$.
	By \eqref{eq:central-block-resonance} and Proposition~\ref{prop:energy-independent-good-parameters}, we can get
	\begin{align*}
		\dist(E,\sigma(\sH_{\mathcal J_{R,N}}))
		\geq\tau_N.
	\end{align*}
	Another application of Proposition~\ref{prop:nonresonant-green-function-decay}
	and the Poisson formula yields
	\begin{equation}\label{eq:10.5}
		|\psi(R)|
		\leq2\lambda (|R|+N+2)^2 e^{-g_\lambda N/2}.
	\end{equation}
	Since $N\geq(|R|-3)/4$, we obtain
	\begin{align*}
		|\psi(R)|
		\leq C(\lambda)e^{-g_\lambda|R|/40}.
	\end{align*}
	Enlarging $C$ to cover the remaining points
	gives \eqref{eq:eigenfunction-exponential-decay}, with
	\begin{align*}
		m_\lambda=\frac{g_\lambda}{40}.
	\end{align*}
	This completes the proof of Theorem \ref{thm:all-coupling-localization}.
\end{proof}

	\section*{Acknowledgments}
This work was  partially supported by NSFC  (12522110).
\section*{Data Availability}
The manuscript has no associated data.
\section*{Declarations}
{\bf Conflicts of interest} \ The authors  state  that there is no conflict of interest.
\section*{AI statement}
The authors declare that the main ideas of this paper are their own. ChatGPT 6.1 sol
(OpenAI) was used to assist with literature searches, language editing, checking calculations
and mathematical arguments. The authors reviewed and verified all AI-assisted material included in the paper and take full responsibility for the accuracy, originality,
and integrity of the entire paper.

\newpage
\appendix{}

\section{Auxiliary estimates for factors}\label{app:determinant-factor-estimates}

\subsection{Proof of Lemma~\ref{lem:van-der-corput-differencing}}

\begin{proof}
Average $H$ translates of the same sum before squaring. This keeps
the differences below $H$, where the arithmetic bounds are useful.
Extend $z_n$ by zero, put $S=\sum_{n\in\Z}z_n$, and observe that
\begin{align*}
HS=\sum_{m\in\Z}\sum_{r=0}^{H-1}z_{m+r}.
\end{align*}
Let $X_m=\sum_{r=0}^{H-1}z_{m+r}$. At most $N+H-1\leq2N$ of
these sums are nonzero. Cauchy-Schwarz inequality gives
\begin{align*}
H^2|S|^2
=\left|\sum_{m\in\Z}X_m\right|^2\leq\#\{m\in\Z:X_m\ne0\}\sum_{m\in\Z}|X_m|^2
\leq2N\sum_{m\in\Z}\left|\sum_{r=0}^{H-1}z_{m+r}\right|^2.
\end{align*}
Expanding the square and grouping the terms by $h=r-s$ yields
\begin{align*}
\sum_{m\in\Z}\left|\sum_{r=0}^{H-1}z_{m+r}\right|^2
&=H\sum_{n\in\Z}|z_n|^2
+2\operatorname{Re}\sum_{h=1}^{H-1}(H-h)
\sum_{n=0}^{N-h-1}z_{n+h}\overline{z_n}.
\end{align*}
Since $\sum_{n\in\Z}|z_n|^2\leq N$ and $H-h\leq H$, we have
\begin{align*}
H^2|S|^2
\leq2N\left[HN+
2H\sum_{h=1}^{H-1}
\left|\sum_{n=0}^{N-h-1}z_{n+h}\overline{z_n}\right|\right].
\end{align*}
Dividing by $H^2N^2$ proves \eqref{eq:van-der-corput-bound}. We have completed the proof.
\end{proof}

\subsection{The discrepancy bound for bounded-variation functions}\label{app:bounded-variation-discrepancy}

\begin{proof}[Proof of \eqref{eq:bounded-variation-discrepancy}]
	Write $\omega=\mu-dt$ and
	$F(t)=\omega([0,t))$. Then
	\begin{align*}
		F(0)=F(1)=0,\
		\|F\|_\infty\leq\mathfrak D(\mu).
	\end{align*}
	For smooth periodic $f$, integration by parts gives
	\begin{align*}
		\int_{\T}f\,d\omega
		=-\int_0^1F(t)f'(t)\,dt,
	\end{align*}
	and hence we otain
	\begin{align*}
			\left|\int_{\T}f\,d\omega\right|
		\leq\mathfrak D(\mu)\Var_{\T}(f).
	\end{align*}
The result follows by uniform approximation with smooth
periodic convolutions, whose variations do not exceed
$\Var_{\T}(f)$. We have concluded the proof.
\end{proof}

\subsection{Mean and variation of the logarithmic kernel}\label{app:logarithmic-kernel}

Use the logarithmic kernel $\ell$ and its truncation $\ell_h$
from Lemma~\ref{lem:circle-product-bounds}. Since $\ell\in L^1(\T)$, its mean
can be computed as follows. Let
$L=\int_0^{\pi/2}\log\sin x\,dx$. Then
\begin{align*}
	2L
	&=\int_0^{\pi/2}\log(\sin x\cos x)\,dx\\
	&=-\frac{\pi}{2}\log2
	+\frac12\int_0^\pi\log\sin u\,du\\
	&=-\frac{\pi}{2}\log2+L.
\end{align*}
Thus $L=-\frac{\pi}{2}\log2$, and consequently
\begin{align*}
	\int_{\T}\ell(t)\,dt=0.
\end{align*}
Let $a_h=\pi^{-1}\arcsin(h/2)$. By symmetry and the
definition of $\ell_h$, we have
\begin{align*}
	\Var_{\T}(\ell_h)
	&=2(\log2-\log h)
	=2\log(2/h),\\
	\int_{\T}\ell_h(t)\,dt
	&=2\int_0^{a_h}\log\frac{h}{2\sin\pi t}\,dt\\
	&\leq2\int_0^{h/4}\log\frac{h}{4t}\,dt
	=\frac h2\leq h,
\end{align*}
where we used $2\sin\pi t\geq4t$ on $[0,1/2]$,
which also implies $a_h\leq h/4$.
\subsection{Proof of Lemma~\ref{lem:free-resolvent-decay}}

\begin{proof}
We use the Combes-Thomas argument \cite{CT73}.
Fix $m\in\Z$ and set
\begin{align*}
\varphi_T(n)=\min\{|n-m|,T\},
\ (\sW_{\mu,T}u)(n)=e^{\mu\varphi_T(n)}u(n).
\end{align*}
Since $|\varphi_T(n+1)-\varphi_T(n)|\leq1$, the Schur test gives
\begin{align*}
\|\sW_{\mu,T}\lambda\Delta\sW_{\mu,T}^{-1}-\lambda\Delta\|
\leq2\lambda(e^\mu-1).
\end{align*}
With $\mu=\mu_\lambda$, the right-hand side is $1/2$. Write
\begin{align*}
\sB=\sW_{\mu_\lambda,T}\lambda\Delta\sW_{\mu_\lambda,T}^{-1}
-\lambda\Delta.
\end{align*}
Since $\|\sR_E\|\leq1$, we have $\|\sB\sR_E\|\leq1/2$.
The Neumann series yields
\begin{align*}
\|\sW_{\mu_\lambda,T}\sR_E\sW_{\mu_\lambda,T}^{-1}\|
=\|\sR_E(\sI+\sB\sR_E)^{-1}\|\leq2.
\end{align*}
Since $\varphi_T(m)=0$, the $(n,m)$ matrix entry satisfies
\begin{align*}
e^{\mu_\lambda\varphi_T(n)}|\sR_E(n,m)|\leq2.
\end{align*}
Taking $T\geq|n-m|$ proves \eqref{eq:free-resolvent-kernel-bound}.

Translation invariance of $\lambda\Delta$ makes $\sR_E$ and $\sC_E$
convolution operators. The identity in \eqref{eq:infinite-volume-cayley-transform} gives
\begin{align*}
|c_E(r)|=2|\sR_E(r,0)|\leq4e^{-\mu_\lambda|r|},
\ r\ne0.
\end{align*}
We complete the proof.
\end{proof}

\subsection{Resolvent compression and trace norms}\label{app:resolvent-compression-and-trace-norms}

\begin{proof}[Proof of Lemma~\ref{lem:cayley-boundary-comparison}]
	Use $\sP$ and $\sQ$ from \eqref{eq:resolvent-compression-identity}, and set
	$\sA=\sR_E^{-1}$. Compressing $\sA\sR_E=\sI$ gives
	\begin{align*}
		\sP
		&=\sP\sA\sR_E\sP\\
		&=(\sP\sA\sP)(\sP\sR_E\sP)
		+\sP\lambda\Delta\sQ\sR_E\sP,
	\end{align*}
	where $\sP\sA\sQ=\sP\lambda\Delta\sQ$ since $\sP\sQ=0$.
	Multiplying on the left by $\sR_{E,N}$ yields
	\eqref{eq:resolvent-compression-identity}. Moreover,
	\begin{align*}
	\sP\Delta\sQ u
	=u(-1)\delta_0+u(N)\delta_{N-1},
	\end{align*}
	so
	\begin{align*}
	\operatorname{rank}(\sR_{E,N}-\sP\sR_E\sP)\leq2.
	\end{align*}
	Since both resolvents have norm at most one,
	\begin{align*}
	\|\sR_{E,N}-\sP\sR_E\sP\|_{\mathfrak S_1}
	\leq
	2\|\sR_{E,N}-\sP\sR_E\sP\|
	\leq4.
	\end{align*}
	The Cayley transform $\sC_E=\sI-2\im\sR_E$ therefore gives
	\eqref{eq:cayley-compression-trace-norm}.
	
	For operators $\sX,\sY$ of norm at most one, the identity
	\begin{align*}
		\sX^k-\sY^k
		=\sum_{j=0}^{k-1}
		\sX^{k-1-j}(\sX-\sY)\sY^j
	\end{align*}
	and the trace-ideal inequality $\|\sA\sB\sC\|_{\mathfrak{S}_1}\le \|\sA\|\cdot\|\sB\|_{\mathfrak{S}_1}\cdot\|\sC\|$ and $\Tr(\sT)\|\le\|\sT\|_{\mathfrak{S}_1}$ yield
	\begin{align*}
	|\Tr(\sX^k-\sY^k)|
	\leq k\|\sX-\sY\|_{\mathfrak S_1}.
	\end{align*}
	Taking
	\begin{align*}
	\sX=\sU_{E,N},
	\ \sY=(\sP\sC_E\sP)\sD_N,
	\end{align*}
	both have norm at most one. Moreover, we can get
	\begin{align*}
	\|\sX-\sY\|_{\mathfrak S_1}
	=\|\sC_{E,N}-\sP\sC_E\sP\|_{\mathfrak S_1}
	\leq8,
	\end{align*}
	since right multiplication by the unitary $\sD_N$
	preserves the trace norm. This proves \eqref{eq:compressed-cayley-trace-error}. The proof is hereby completed.
\end{proof}
\subsection{Comparison with the path sum}\label{app:finite-trace-path-comparison}

\begin{proof}[Proof of \eqref{eq:finite-trace-path-comparison}]
	Let $M=M_N$, $\sP=\sP_N$, and define
	\begin{align*}
		\mathscr Y_N&=(\sP\sC_E\sP)\sD_N,\\
		\mathscr Y_{N,M}&=(\sP\sC_E^{[M]}\sP)\sD_N,\\
		\sW_M&=\sC_E^{[M]}\sD_{\alpha,\beta}.
	\end{align*}
	By \eqref{eq:compressed-cayley-trace-error}, we have
	\begin{align*}
	|\Tr(\sU_{E,N}^k)-\Tr(\mathscr Y_N^k)|\leq8k.
	\end{align*}
	Moreover, \eqref{eq:cayley-truncation-bounds} and
	$\|T\|_{\mathfrak S_1}\leq N\|T\|$ give
	\begin{align*}
	\|\mathscr Y_N-\mathscr Y_{N,M}\|_{\mathfrak S_1}
	\leq N\varepsilon_N.
	\end{align*}
	Since  $\|\sY_N\|,\|\sY_{N,M}\|\le 1+\varepsilon_N$,
		\begin{align*}
		\sY_N^k-\sY_{N,M}^k
		=\sum_{j=0}^{k-1}
		\sY_N^{k-1-j}(\sY_N-\sY_{N,M})\sY_{N,M}^j
	\end{align*}
	yields
	\begin{align*}
	|\Tr(\mathscr Y_N^k)-\Tr(\mathscr Y_{N,M}^k)|
	\leq Nk\varepsilon_N(1+\varepsilon_N)^{k-1}
	\leq2Nk\varepsilon_N.
	\end{align*}
	
The two diagonal entries coincide outside the boundary
layers of total size at most $2kM$. Hence, we can get
\begin{align*}
	\left|\Tr(\mathscr Y_{N,M}^k)
	-\sum_{n=0}^{N-1}\sW_M^k(n,n)\right|\leq 2kM\cdot 4=8kM.
\end{align*}
Combining the three estimates and using \eqref{eq:closed-path-diagonal-entry}
gives \eqref{eq:finite-trace-path-comparison}. The proof is hereby completed.
\end{proof}

\section{Spectral separation and infinite-volume details}\label{app:phase-root-and-block-spectral-separation}

\subsection{The quantitative bound in Lemma~\ref{lem:spectral-gap-to-phase-root-separation}}\label{app:phase-root-separation-proof}

\begin{proof}[Proof of Lemma~\ref{lem:spectral-gap-to-phase-root-separation}]
	Since $\|\Delta_N\|\leq2$ and $|E|\leq E_*$,
	\begin{align*}
	\|\sK_{E,N}-\im\sI_N\|
	\leq2\lambda+E_*+1=C.
	\end{align*}
	By \eqref{eq:cleared-matrix-kernel-identity} and $\|v\|=1$, we have
	\begin{align*}
	2\leq C\|u_0\|,
	\ \|u\|\geq\|u_0\|-|w-w_0|.
	\end{align*}
	Thus, if $|w-w_0|\leq C^{-1}$, then we get
	\begin{align*}
	\|u\|\geq C^{-1}.
	\end{align*}
	Using the residual identity and the spectral theorem, we obtain
	\begin{align*}
		\delta\leq\dist(E,\sigma(\sH_N))\leq\frac{\|(\sH_N-E\sI_N)u\|}{\|u\|}\leq C^2|w-w_0|.
	\end{align*}
	If $|w-w_0|>C^{-1}$, then we conclude
	\begin{align*}
	|w-w_0|>C^{-1}\geq\delta C^{-2}.
	\end{align*}
	Hence, in either case, we have
	\begin{align*}
	|w-w_0|\geq C^{-2}\delta.
	\end{align*}
	The argument only requires $\sH_N$ to be defined at $w$,
	so it also applies when $w_0$ is a pole. We close the proof here.
\end{proof}

\subsection{Counting the pole-free intervals in Lemma~\ref{lem:two-block-spectral-separation-measure}}\label{app:shear-pole-count}

\begin{proof}
	Fix $\vartheta$. Since $s_n\neq0$ on $\boI_{c,N}$ and
	$\mathcal J_{R,N}$, the
	pole equation
	\begin{align*}
	s_nt\equiv\tfrac12-\vartheta\pmod1
	\end{align*}
	has exactly $|s_n|$ solutions on $\T$ for each point $n\in\Z$.
	Moreover,
	\begin{align*}
	\#(\boI_{c,N}\cup\mathcal J_{R,N})\leq5N,
	\ |n|\leq7N,
	\end{align*}
	so that
	\begin{align*}
	\sum_{n\in\boI_{c,N}\cup\mathcal J_{R,N}}|s_n|
	\leq5N(49N^2+4N^2)=265N^3.
	\end{align*}
	Thus the pole-free part of $\T$ consists of at most
	$265N^3+1$ intervals.
	
Let
\begin{align*}
	\sigma_1(t)=\sigma(\sH_{\boI_{c,N}}(\vartheta,t)),
	\ \sigma_2(t)=\sigma(\sH_{\mathcal J_{R,N}}(\vartheta,t)).
\end{align*}
By the derivative bound, the set of parameters for which
any fixed pair of eigenvalues differs by less than $\delta$
has measure at most
	\begin{align*}
	\mes\{t:\ |\zeta_k(t)-\eta_j(t)|<\delta\}
	\leq\frac{\delta}{3\pi N^2}.
\end{align*}
Consequently, using $L/N\leq5/2$ and $N\geq2$, we have
\begin{align*}
	\mes\{t:\ \dist(\sigma_1(t),\sigma_2(t))<\delta\}
	&\leq(265N^3+1)\frac{L^2\delta}{3\pi N^2}\\
	&\leq\frac{53025}{96\pi}N^3\delta\\
	&<200N^3\delta.
\end{align*}
The same bound holds after removing $\boP$.
Integrating over $\vartheta$ and using the measure
preservation of \eqref{eq:parameter-shear} proves
\eqref{eq:two-block-spectral-gap-measure}. This completes the proof.
\end{proof}

\newcommand{\etalchar}[1]{$^{#1}$}

\end{document}